\documentclass[nofootinbib,amsfonts,prd,aps]{revtex4}
\usepackage{graphicx}
\begin{document}

\title{A scenario for black hole evaporation on a quantum geometry}

\author{Rodolfo Gambini$^{1}$,
Jorge Pullin$^{2}$}
\affiliation {
1. Instituto de F\'{\i}sica, Facultad de Ciencias, 
Igu\'a 4225, esq. Mataojo, 11400 Montevideo, Uruguay. \\
2. Department of Physics and Astronomy, Louisiana State University,
Baton Rouge, LA 70803-4001}

\begin{abstract}
  We incorporate elements of the recently discovered exact solutions
  of the quantum constraints of loop quantum gravity for vacuum
  spherically symmetric space-times into the paradigm of black hole
  evaporation due to Ashtekar and Bojowald.  The quantization of the
  area of the surfaces of symmetry of the solutions implies that the
  number of nice slices that can be fit inside the black hole is
  finite. The foliation eventually moves through the region where the
  singularity in the classical theory used to be and all the particles
  that fell into the black hole due to Hawking radiation emerge
  finally as a white hole.  This yields a variant of a scenario
  advocated by Arkani-Hamed {\em et al.}  Fluctuations in the horizon
  that naturally arise in the quantum space time allow radiation to
  emerge during the evaporation process due to stimulated emission
  allowing evaporation to proceed beyond Page time without reaching
  the maximum entanglement limit until the formation of the white
  hole. No firewalls nor remnants arise in this scenario.
\end{abstract}

\maketitle

In a recent development, the exact space of solutions of the equations
of loop quantum gravity for vacuum, spherically symmetric space-times
was found \cite{spherical}. This was due to the realization that a
rescaling and linear combination of the constraints of canonical
quantum gravity in the spherical case yields a constraint algebra that
is a Lie algebra. This allows to complete the Dirac quantization
procedure for the model. Remarkably, the space of physical states can
be found in closed form. It is based on one dimensional spin
networks. The proximity of the nodes of the spin networks is bounded below
by the condition of the quantization of the areas of the spheres of
symmetry. The singularity that is inside black holes in classical
general relativity is replaced by a region where a description in
terms of a semiclassical geometry is not possible and through it one
tunnels into another region of space-time. Although this solution
corresponds to an eternal time independent black hole, elements from
its description can be heuristically incorporated into black hole
evaporation scenarios. We will see that they can have interesting
implications in those situations.

A paradigm for black hole evaporation in loop quantum gravity was put
forward by Ashtekar and Bojowald \cite{asbo}. Motivated by loop
quantum gravity analyses of the black hole interior treating it as a
Kantowski--Sachs cosmology \cite{ks} and of results in CGHS $1+1$
dimensional model \cite{cghs}, the picture that emerges is the one
depicted in  1. As a star collapses to form a black hole, a
trapping horizon forms, initially space-like near $r=0$ and eventually
becomes time-like, ending up by touching the region where a semiclassical
space-time description is not available.
\begin{figure}[h]
\includegraphics[height=5.5cm]{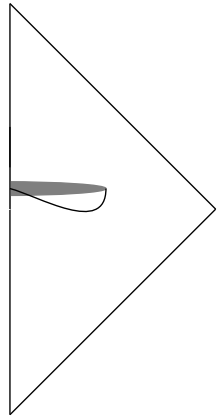}
\caption{The Penrose diagram of the Ashtekar--Bojowald paradigm. As
  matter collapses, a trapping horizon forms, initially spatial and
  eventually timelike. The shaded region is not well approximated by a
  semiclassical space-time and is where the classical singularity
  was.}
\end{figure}

We would like to argue that when one incorporates elements of the new
exact solution, namely, the spatial discreteness implied by the
quantization of area in loop quantum gravity and the fact that states
are superpositions with different masses and spin networks, to the
above paradigm, a new picture emerges that is compatible with
Susskind's black hole complementarity \cite{susskind}. The three
principles of black hole complementarity are: 1) That the evolution is
unitary, there is an S matrix describing the evolution of infalling
matter to Hawking radiation. 2) Outside the horizon, physics can be
described by quantum field theory on a classical background with small
but important quantum gravity corrections.  In our
case the background is represented by a geometry given by the
expectation values of the metric on suitable quantum states that
approximate the classical behavior. 3) To a distant observer a black
hole appears as a quantum system with discrete energy levels for the
purposes of entropy calculations. In our
case, this is implied by the quantization of the areas, as we will
note later. In addition it is usually assumed that an observer falling
into the black hole does not notice anything unusual when crossing the
horizon \cite{amps}.

A standard way to describe Hawking radiation is to consider the
construction of ``nice slices'' \cite{mathur,nima}. These are spatial
slices of the space-time that penetrate the horizon interpolating
between $t={\rm const.}$ slices in the exterior to $r={\rm const.}$
slices in the interior (in Schwarzschild coordinates). This requires a
stretching in the region around the horizon. The stretching implies
that the vacuum on one slice is not the vacuum on future slices and
this implies creation of pairs of particles and accounts for the
Hawking radiation. Each particle emitted in the exterior becomes
entangled with a particle inside the black hole. For an observer in
the asymptotic region, the radiation appears thermal, since part of
the information is lost, at least locally, and one needs to trace over
the degrees of freedom not accessible behind the horizon. The slices
are usually constructed with a spacing in the exterior of the black
hole of the order of twice the Schwarzschild radius. This ensures that
the radiation produced has the energy corresponding to the maximum of
the standard spectrum of Hawking radiation. 
\begin{figure}[h]
\includegraphics[height=5.5cm]{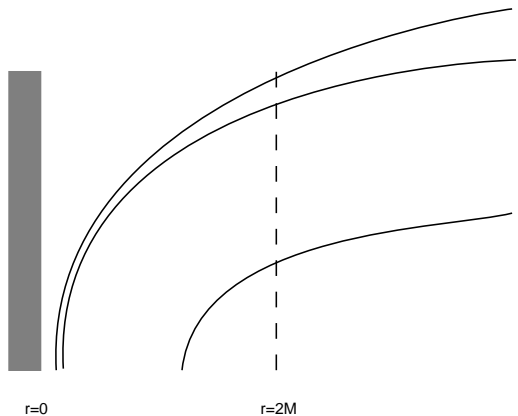}
\caption{The ``nice slices'' that enter the black hole and whose
  stretching is responsible for the Hawking radiation. Outside they
  asymptote to surfaces $t={\rm const.}$ and inside to $r={\rm
    const.}$ in ordinary Schwarzschild coordinates.} 
\end{figure}

One of the new elements introduced by the quantum space-time of loop
quantum gravity is that the separation of the slices is bounded below
by the rule of the quantization of the area of the spheres of
spherical symmetry.  The background quantum space-time that emerges
from the new exact solution for the physical space of states in
spherical symmetry is based on one dimensional spin networks. One has
a graph $g$ on which the network is based and a set of integers
$\vec{k}$ that are the valences the edges of the spin network. The
values that the areas of the spheres of symmetry can take on these
states are $4\pi \ell_{\rm Planck}^2 k$ at the various links of the
spin network.  This implies that for the slices in the interior of the
black hole that are close to the singularity one has a separation in
the radial coordinate of order $\ell_{\rm Planck}$. Elsewhere, the
separation of the slices is bounded below by $\ell_{\rm
  Planck}^2/(2r)$. The value $k=0$ has to be eliminated in order to
have a self-adjoint metric operator and that implies that the
singularity present in the classical Schwarzschild solution is
eliminated in this approach.  This in turn means that in the interior
of the black hole can only accommodate a finite number of slices
bounded above by ${\rm Int}((2 G M)^2/\ell^2_{\rm Planck})$.

Remarkably, this simple one dimensional model of black hole
reproduces correctly the entropy calculation. If one considers all
possible states in the interior, its number is bounded by $2^{(2 G
  M)^2/\ell^2_{\rm Planck}}$ (this is the number of all possible
vectors $\vec{k}$ that can be assigned to the black hole interior for
a given mass $M$). Its logarithm then yields an entropy proportional to the
area in Planck units with a factor of order unity.

In turn, this means that from the point of view of an observer far
away from the black hole, the traditional evaporation process lasts a
time ${\rm Int}((2 G M)^2/\ell^2_{\rm Planck}) (4 G M)/c$. This gives
$16 G^3 M^3/(\hbar G c^4)$, which is of the order of magnitude of the
Page time. At this time, the nice slices reach the region 
where the classical singularity used to be and quantum field
theory in curved space-time is not a good approximation anymore to the
relevant physics. Once the slices traverse the region where the
singularity was in the classical theory, all the particles that fell
in due to Hawking radiation during the evaporation emerge in a white
hole.  Notice that the previous evaporation time is only 
an order of magnitude estimate, we are ignoring several important
effects, including the black hole evaporation in its calculation. 

After the slices cross the region where the classical
singularity used to be and become more and more like ordinary slices
of Minkowski space, the usual Hawking evaporation switches off and the
particles that fell into the black hole are allowed to emerge. At this
point we cannot give a precise picture of how the radiation that
emerges after the traditional evaporation stops behaves in
detail. That would require studying how the particles that fell into
the black hole (and are entangled with the earlier outgoing radiation)
interact with the region of high curvature surrounding where the
classical singularity used to be. A preliminary study of shells
crossing the region, exhibiting some of the relevant effects can be
seen in \cite{shells}. Once the nice slices went through
that region, they have it in their past and therefore even their
discussion requires a full quantum treatment.

An objection that could be raised to this scenario is that as the
black hole evaporates, it is well know that the interior becomes more
and more entangled with the outgoing radiation until there are not
enough degrees of freedom left in the interior to continue the
evaporation \cite{page}. Therefore Hawking evaporation should stop in
a regime where quantum field theory in curved space time should work
very well.  This, among other reasons, led to the proposal of
firewalls to avoid this problem \cite{amps}. It would also imply that
the scenario we are proposing could not take place. This point of view
does not take into account that in a situation with a quantum
geometry, one will have fluctuations of the position of the
horizon. In the exact solution presented in \cite{spherical} this is
taken into account since the mass of the black hole is a Dirac
observable with continuous spectrum and the generic solution
representing a classical space time will be given by a superposition
of states with different values of the mass, centered in a given
classical value.  That means that the quantum state near the horizon
will fluctuate from being a vacuum into having particles.
Specifically, every time a particle of the Hawking radiation is
emitted, the mass of the black hole will fluctuate by an amount given
by the time-energy uncertainty relation. The characteristic time for
emission is given by the separation of the ``nice slices'' and is
therefore of the order of $2GM/c$. The horizon therefore carries out a
``random walk'' in position and after a Page time the fluctuation in
its position is of the order of the Planck mass. This implies a
separation in the distance between the inner and outer extrema of the
fluctuations of the order of $2\int_{r_S}^{r_S+\ell_{\rm Planck}}
\sqrt{1-r_S/r}^{-1} dr=2\sqrt{2 G M \ell_{\rm Planck}}+O(\ell_{\rm
  Planck}^{3/2})$. So it is a cumulative effect on the position of
the horizon due to the back-reaction of the particle production. The
large differences in characteristic frequencies associated with the
positions of the horizon in the fluctuation ---due to the high
blueshifts close to the horizon---, imply that the vacuum in one
position will be significantly different than the vacuum in the
other. This would lead to stimulated emission in the Hawking radiation
of similar nature to that due to thermal and other types of
fluctuations of the horizon in black hole analogue systems
\cite{jacobson}. A heuristic picture is that a particle created by
pair production for a horizon will stimulate emission of particles
with the same quantum numbers for a horizon further out as the
position of the horizon fluctuates due to quantum effects.  This extra
emission allows the black hole to lose mass without
increasing the entanglement entropy.  Stimulated emission (with a very
different origin) has been proposed \cite{adami} as a solution of the
information problem by itself, however it does not appear be a strong
enough effect to account for the exit of all the information at the
end of the process. Our scenario does not face this problem as the
information can further exit in significant quantities through the
white hole (plus the effect leading to stimulated emission appears
larger).  Hawking radiation will fail to be perfectly thermal due to
stimulated emission. This will manifest itself in fluctuations in the
temperature. The radiation may also exhibit a certain degree of
coherence or non-thermality due to the stimulated emission, absent in
ordinary Hawking radiation. In other words, entanglement grows at a
smaller rate than the one predicted by ordinary Hawking
radiation. These deviations could be calculated in detail and the
magnitude of this effect estimated with the quantum space-time
techniques of \cite{hawkingus} but they would require certain
modifications of how one treats the matter Hamiltonian, that exceed
the scope of this note. So at this point we conjecture that the
combined effect of spontaneous and stimulated emission will allow the
evaporation to continue till the white hole is formed but this remains
to be confirmed.

This mechanism of finite number of nice slices therefore provides a
variant of an implementation of a proposal by Arkani-Hamed {\rm et
  al.}  \cite{nima}. In their original proposal the limitation was
given by bounds on the accuracy for measuring times between two
successive slices that was heuristic. Here such limitation arises from
the loop quantization of the space-time and the traditional
evaporation switches off gradually after the slices make it through the
region where the classical singularity used to be. 

\begin{figure}[h]
\includegraphics[height=5.5cm]{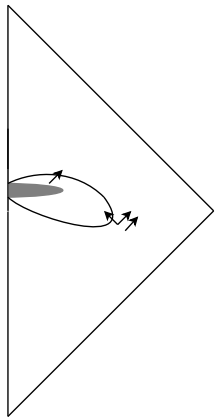}
\caption{The proposed paradigm. A trapping horizon forms during the
  Hawking radiation period. When the radiation ends one has a white
  hole that radiates the information that had fallen into the trapping
  horizon in the process of generation of Hawking radiation and no
  remnant is left behind. The arrows indicate particles of the Hawking
  radiation outgoing and infalling into the black hole and particles
  of the white hole exiting.}
\end{figure}

Since in our approach there is no singularity, the particles of the
Hawking radiation that fell into the black hole can in principle
traverse the region where the classical singularity used to be. The
detailed description of that traversing is involved, but in principle
there is no reason for the information that fell into the black hole
not to get out by traversing the region where the singularity used
to be. The resulting picture is that of a white hole, with
considerable similarity to the original white hole proposal of Hawking
\cite{hawking}: ``To an external observer a white hole is
indistinguishable from a black hole. The process of hole formation and
evaporation is completely time-symmetric. The irreversibility which
arises in the classical limit is just a statistical effect''. In the
case we are considering there is no reason for the process to be
time-symmetric, though. 

As the picture that emerges is that of a white hole, it suggests that
no ``remnant'' is produced, although the true details will depend on
the dynamics in the region of high curvature which at the moment we do
not have. Asymptotically, the information could therefore potentially
be retrieved. The information of the first particle of each produced
pair could be retrieved at infinity through the usual Hawking
radiation (and the stimulated emission) and the information of the
particle that fell into the black hole is radiated to infinity in the
white hole phase. What we are doing in our proposal is to start with a
black hole that morphs into a white hole.  However, since the
particles that emerge through the white hole had complicated
interactions in the region of high curvature, the radiation of the
white hole is not the same as the one produced during the black hole
phase. This has some similarities with the model considered by Parikh
and Wilczek \cite{parikh} in that there is a region of space-time that
is in causal contact with a region where one needs the full quantum
theory for its description and that is the source of the asymmetry in
the radiation. In particular in the white hole phase, it is unlikely
to have a purely thermal spectrum.  Notice that some of the arguments
given against white holes that argue they are unstable are based on
the existence of Cauchy horizons for the space-time, which in the
Ashtekar-Bojowald picture do not emerge.

Notice that we have respected complementarity, there was no need to
invoke firewalls \cite{amps} and the information is naturally
retrieved completely without remnants. However, although we have a
scenario where quantum field theory in curved spacetime applies
locally in the regions of low curvature, our construction requires
slices that went through a region where there is not a classical
space-time description which will modify usual predictions in the
final stages of evaporation.  In particular traditional Hawking
radiation switches off and is replaced by emission of
particles that had fallen into the black hole. An important caveat is
that our calculations are based on spherical symmetry. It is
reassuring that remarkably, the entropy calculation works out (at an
order-of-magnitude level) and that the bound on the number of slices
can be arrived at from different perspectives, that are also valid
without spherical symmetry.  Loop quantum gravity calculations of
Hawking radiation are now available \cite{hawkingus} and in particular
imply the elimination of singularities in physical quantities, like
the expectation value of the stress--energy tensor. This opens up the
possibility of carrying out back reaction calculations. With this, it
will be possible relatively soon to further explore the presented
paradigm with detailed calculations.

We wish to thank Don Marolf for many helpful discussions.
This work was supported in part by grant NSF-PHY-0968871, funds of the
Hearne Institute for Theoretical Physics, CCT-LSU and Pedeciba.

\end{document}